\UseRawInputEncoding

\documentclass[floatfix,%
 reprint,
 superscriptaddress, % <--Group the authors
 amsmath,amssymb,
 aps,
 pra,
]{revtex4-2}

\usepackage{natbib}
\usepackage{placeins}
\usepackage{textcomp} 
\usepackage{graphicx}% Include figure files
\usepackage{dcolumn}% Align table columns on decimal point
\usepackage{bm}% bold math
\usepackage{xcolor} % font color
\usepackage{jabbrv}
\definecolor{PRBlue}{RGB}{65, 105, 225} % navy blue
\usepackage[colorlinks=true,
            linkcolor=PRBlue,
            citecolor=PRBlue,
            urlcolor=PRBlue]{hyperref}
\makeatletter
\def\section{%
  \@startsection{section}% 
  {1}% 
  {0pt}% 
  {2.0ex plus 0.8ex minus .3ex}% ）
  {1.2ex plus 0.3ex}
  {\normalfont\large\bfseries\raggedright}
}

\def\subsection{%
  \@startsection{subsection}
  {2}
  {0pt}
  {1.5ex plus 0.5ex minus .2ex}
  {0.8ex plus .2ex}
  {\normalfont\normalsize\bfseries\raggedright}
}
\makeatother

\begin{document}

\preprint{APS/123-QED}

\title{Thermally accessible broadband soliton microcombs in silicon carbide enabled by dynamic polarization control}% Force line breaks with \\

% \author{Andreas Jacobsen\,\textsuperscript{1,*}, Yang Liu\,\textsuperscript{1,*}, Thibault Wildi\,\textsuperscript{2,*}, Yanjing Zhao\,\textsuperscript{1}, Chaochao Ye\,\textsuperscript{1}, Yi Zheng\,\textsuperscript{1}, Camiel Op de Beeck\,\textsuperscript{3}, José Carreira\,\textsuperscript{3}, Kresten Yvind\,\textsuperscript{1}, Michael Geiselmann\,\textsuperscript{3}, Tobias Herr\,\textsuperscript{2,\dag}, Minhao Pu\,\textsuperscript{1,\ddag}}

% \affiliation{DTU Electro, Department of Electrical and Photonics Engineering, Technical University of Denmark, 2800 Kongens Lyngby, Denmark}
% \affiliation{Deutsches Elektronen-Synchrotron DESY, Notkestr. 85, 22607 Hamburg, Germany}
% \affiliation{LIGENTEC SA, EPFL Innovation Park, 1024 Ecublens, Switzerland}

\author{Haoyang Tan}
\thanks{These authors contributed equally.}
\affiliation{DTU Electro, Department of Electrical and Photonics Engineering, Technical University of Denmark, 2800 Kongens Lyngby, Denmark}

\author{Yi Zheng}
\thanks{These authors contributed equally.}
\affiliation{DTU Electro, Department of Electrical and Photonics Engineering, Technical University of Denmark, 2800 Kongens Lyngby, Denmark}

\author{Xiyuan Lu}
\affiliation{Microsystems and Nanotechnology Division, Physical Measurement Laboratory, National Institute of Standards and Technology, Gaithersburg, MD, USA}
\affiliation{Joint Quantum Institute, NIST/University of Maryland, College Park, MD, USA}

\author{Yang Liu}
\affiliation{DTU Electro, Department of Electrical and Photonics Engineering, Technical University of Denmark, 2800 Kongens Lyngby, Denmark}

\author{Andreas Jacobsen}
\affiliation{DTU Electro, Department of Electrical and Photonics Engineering, Technical University of Denmark, 2800 Kongens Lyngby, Denmark}

\author{Kresten Yvind}
\affiliation{DTU Electro, Department of Electrical and Photonics Engineering, Technical University of Denmark, 2800 Kongens Lyngby, Denmark}

\author{Kartik Srinivasan}
\affiliation{Microsystems and Nanotechnology Division, Physical Measurement Laboratory, National Institute of Standards and Technology, Gaithersburg, MD, USA}
\affiliation{Joint Quantum Institute, NIST/University of Maryland, College Park, MD, USA}

\author{Minhao Pu}
\thanks{mipu@dtu.dk}
\affiliation{DTU Electro, Department of Electrical and Photonics Engineering, Technical University of Denmark, 2800 Kongens Lyngby, Denmark}

\date{\today}% It is always \today, today,

\begin{abstract}
% [Original version] Optical microcombs based on high-Q microresonators are promising chip-scale light sources for applications ranging from optical communication to spectroscopy and metrology. However, thermo-optic instabilities pose a major challenge for stable soliton comb access. Self-cooling via auxiliary modes can stabilize the intracavity power, yet the diverted cooling power is not utilized for soliton generation, limiting comb power and bandwidth. Here, we propose a thermal compensation scheme through dynamic polarization control. In this scheme, a controlled fraction of the pump is coupled to the TM mode to provide effective self-cooling and ensure reliable soliton generation. After soliton formation, polarization rotation and pump tuning redirect the cooling power to the TE comb mode, enabling single-soliton generation with enhanced power and bandwidth. This dynamic polarization-based thermal compensation provides thermally stabilized soliton initiation and improved microcomb performance. 

Optical microcombs generated in high-Q microresonators are promising chip-scale light sources for applications ranging from optical communications to spectroscopy and metrology. However, thermo-optic instabilities remain a major obstacle to reliable soliton access. Self-cooling using auxiliary modes can stabilize the intracavity power, yet part of the power is continuously allocated to thermal compensation rather than comb generation, thereby limiting comb power and bandwidth. Here we propose a thermal compensation scheme based on dynamic polarization control. During soliton initiation, a fraction of the pump is coupled to an orthogonally polarized mode to provide self-cooling and ensure reliable soliton access. After soliton formation, polarization rotation and pump tuning transfer this cooling power to the comb-generating mode, enabling efficient single-soliton operation. Using this approach, we experimentally demonstrate a broadband 108-GHz-FSR single-soliton microcomb spanning over 450 nm, together with approximately 39\% improvement in the 20-dB bandwidth and 60\% increase in comb power relative to the static self-cooling configuration. This dynamic polarization-based thermal compensation enables efficient use of available laser power and provides a practical route to high-performance soliton microcombs in platforms with strong thermo-optic effects.
\end{abstract}


\maketitle
\let\thefootnote\relax
\footnotetext{* These authors contributed equally}
\footnotetext{\dag\ tobias.herr@desy.de}
\footnotetext{\ddag\ mipu@dtu.dk}

%\tableofcontents

\section{Introduction}

Optical frequency combs (OFCs) are important light sources consisting of equally spaced and phase-coherent spectral lines. In recent years, chip-based OFCs (microcombs) generated via parametric four-wave mixing (FWM) in high-quality microresonators have become a promising alternative to conventional mode-locked laser combs, featuring a compact footprint, low power consumption, and compatibility with chip-scale photonic integration \cite{Kippenberg2011Microresonator-basedCombs,Yang2024EfficientCombs}. They have been widely explored in coherent communication \cite{Pfeifle2014CoherentCombs,Marin-Palomo2017Microresonator-basedCommunications}, dual-comb spectroscopy \cite{Suh2016MicroresonatorSpectroscopy,Yu2018Silicon-chip-basedSpectroscopy}, imaging \cite{Bao2019MicroresonatorImaging,Marchand2021SolitonTomography}, and frequency metrology \cite{Trocha2018UltrafastCombs,Riemensberger2020MassivelyMicrocomb}. 
   
    Over the past two decades, substantial progress has been made in advancing high-quality factor (Q) microresonators across a broad range of integrated nonlinear photonic platforms \cite{Gaeta2019Photonic-chip-basedCombs}. Silicon-based material systems including Si, SiO\textsubscript{2}, Hydex, and Si\textsubscript{3}N\textsubscript{4} \cite{Griffith2015Silicon-chipGeneration, DelHaye2007OpticalMicroresonator,Yi2015SolitonMicroresonator, Razzari2010CMOS-compatibleOscillator, Bao2017DirectMicroresonators,Pfeiffer2017Octave-spanningMicroresonators,Brasch2016PhotonicRadiation,Joshi2016ThermallyMicroresonators,Li2017StablyRegime} have been extensively developed, alongside emerging nonlinear platforms such as AlN, LiNbO\textsubscript{3}, GaN, GaP, AlGaAs, and chalcogenide glasses \cite{Liu2021AluminumSelf-referencing, Weng2021DirectlyMicroresonator, He2019Self-startingMicrocomb, Wang2019MonolithicModulation, Gong2020Near-octaveMicrocomb, Zheng2022IntegratedPhotonics, Wilson2020IntegratedPhotonics, Chang2020Ultra-efficientMicroresonators, Pu2016EfficientAlGaAs-on-insulator, Xia2023IntegratedCombs}. Recently, silicon carbide (SiC) has attracted growing interest for integrated nonlinear photonics \cite{Guidry2022QuantumMicrocombs,Cai2022Octave-spanningPlatformb,Wang2022SolitonPlatform} due to its intrinsic $\chi^{(2)}$  \cite{Zheng2025EfficientNanowaveguides} and $\chi^{(3)}$, and relatively large Kerr nonlinear index ($\sim10^{-18}~\mathrm{m^2/W}$) \cite{Zheng20194H-SiCPhotonics,Lukin20204H-silicon-carbide-on-insulatorPhotonics}.  However, the relatively large thermo-optic coefficient ($\approx4.2\times10^{-5}~\mathrm{K^{-1}}$ for 4H-SiC, roughly twice that of Si\textsubscript{3}N\textsubscript{4})   \cite{Shi2021ThermalPlatforms,Arbabi2013MeasurementsResonances} leads to pronounced resonance shift under high optical power,  which poses challenges for accessing the red-detuned regime required for soliton formation \cite{Li2017StablyRegime}. Effective thermal management is therefore essential for reliable single-soliton generation in SiC microresonators.
     
\begin{figure*}[htpb]
	\begin{center}
		\includegraphics[width=1\linewidth]{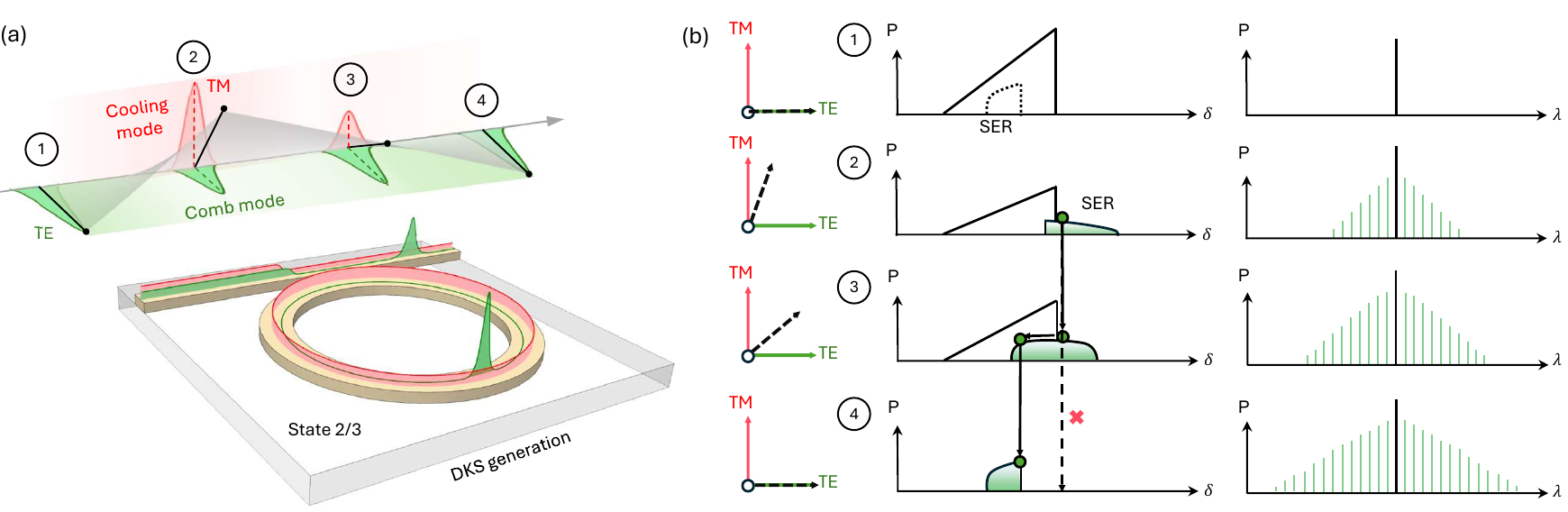}
		\caption{\textbf{Broadband single-soliton comb generation via thermal compensation using dynamic polarization control.} (a) Schematic of pump polarization control. The TE mode is employed for comb generation, while the TM mode provides self thermal cooling. State 1 corresponds to a purely-TE polarized pump. State 2 and 3 represent obliquely polarized pumping conditions, under which a single soliton can be accessed. From State 2 to State 4, the TM fraction is gradually reduced until the input returns to pure TE polarization. (b) Evolution of comb power ($P$) as a function of pump detuning ($\delta$) and corresponding comb spectra across different states. The procedure is initiated at State 2, where an obliquely polarized pump introduces TM-assisted thermal cooling and enables stable access to the soliton existence range (SER, green shaded regime), mitigating the thermo-optic instabilities that typically obscure the soliton state (dashed line) in the purely TE-polarized case (State 1). After soliton initiation, the cooling power is redirected to the TE comb mode via polarization rotation and pump blue-detuning (no speed requirement), enhancing both output comb power and bandwidth (from State 2 to State 4). The comb spectra corresponding to the maximum comb detuning on the right are schematic representations based on experimental observations, illustrating the continuous broadening bandwidth as cooling power is redirected to the comb mode.}
		\label{fig:1}
	\end{center}
\end{figure*}
    
    A variety of strategies have been explored to manage thermo-optic instabilities in microresonators. Rapid tuning of either the pump laser or the resonator temperature \cite{Guo2017UniversalMicroresonators,Xue2016ThermalResonators,Joshi2016ThermallyMicroresonators} as well as the power kicking technique \cite{Yi2016ActiveMicroresonators,Brasch2016BringingState} can be used to transiently bypass the thermal response. Device-level approaches based on dispersion engineering and cavity length design have also been used to facilitate direct access to Kerr soliton states, typically by operating in large free-spectral-range (FSR) regimes \cite{Wu2023AlGaAsTemperature,Jacobsen2025High-powerGeneration}. Auxiliary lasers  have also been employed to stabilize the intracavity power and counteract thermal instabilities \cite{Zhang2019Sub-milliwatt-levelLaser,Zhou2019SolitonMicrocavities,Zhao2021SolitonLaser,Zhao2025ThermalGeneration}, but they often require additional components that complicate system control and hinder full photonic integration. Leveraging these strategies, soliton microcombs have recently been demonstrated in the SiC platform \cite{Guidry2022QuantumMicrocombs, Wang2022SolitonPlatform, Sun2024DirectlyInvited, Zheng2026Octave-SpanningResonators}. However, these implementations typically rely on large-FSR resonators, auxiliary-laser cooling or cryogenic operation, all of which impose device design constraints or increase system complexity. To overcome these limitations, single-pump thermal compensation schemes have been explored in other integrated platforms, offering promising alternatives by simultaneously exciting a comb mode and an auxiliary mode via cross-polarization pumping \cite{Li2017StablyRegime,Lei2022ThermalSelf-cooling,Aldhafeeri2024LowMicrocombs} or dual-mode pumping \cite{Weng2021DirectlyMicroresonator,Weng2022Dual-modeSolitons,Weng2024Turn-keyMicroresonators}. Nevertheless, in most self-cooling configurations, a portion of the input power is diverted to the auxiliary mode for thermal compensation and remains trapped there after soliton formation. This prevents the full available pump power from contributing to comb generation, thereby limiting the achievable comb bandwidth and output power. The impact of this limitation becomes particularly pronounced for broadband and small FSR microcombs, which require both higher soliton-driving power and stronger thermal stabilization \cite{Sun2024DirectlyInvited}.
    
    In this work, we propose and demonstrate a thermal compensation scheme for reliable single-soliton generation in SiC microresonators through dynamic polarization control. An obliquely polarized pump partially couples to the fundamental transverse electric (TE) mode for comb generation and the fundamental transverse magnetic (TM) mode for thermal compensation, enabling soliton initiation. Once solitons are accessed, the cooling power is dynamically redirected to the comb-generating TE mode through controlled polarization rotation and pump wavelength tuning, allowing the previously diverted power to directly enhance the comb power and bandwidth. This approach enables thermally accessible single-soliton generation while fully utilizing the available pump power for soliton comb operation. Using this method, we experimentally demonstrate a 108-GHz-FSR single-soliton comb spanning 450 nm, combining broad spectral coverage with moderate line spacing suitable for applications requiring both wide bandwidth and manageable channel counts.
    
\section{Results}

\subsection{Operation principle}
Fig.~\ref{fig:1} schematically illustrates the principle of single-soliton generation in a microresonator via thermal compensation using an obliquely polarized pump. The subsequent cooling power reuse further enhances the comb power and bandwidth. In this scheme, the TE mode is designated as the comb mode for soliton generation, while a red-detuned TM mode serves as a cooling mode for thermal compensation. Under a purely TE-polarized input (State 1), scanning the pump wavelength from the blue- to the red-detuned side of the cavity resonance triggers a transition from the modulation instability (MI) regime to the soliton regime. This transition is accompanied by a sharp drop in intracavity power, inducing strong thermo-optic instability that shortens the soliton step and hinders deterministic access. By contrast, with an obliquely polarized pump (State 2/3), a controlled fraction of the pump power couples into the TM cooling mode as the pump enters the red-detuned regime. This auxiliary channel mitigates the abrupt power drop between the MI and soliton states, ensuring reliable access to the soliton step (indicated by the green shaded regions). 

Once a single soliton is established, the power previously coupled to the cooling mode can be redirected into the TE comb mode to enhance the comb power and bandwidth. As depicted in Fig.~\ref{fig:1}(b), the SER shrinks and blue-shifts as the TE component increases. Consequently, a direct polarization rotation back to pure TE is inhibited if there is no overlap between the SER of the initial polarization state and that of the pure TE state. To overcome this, an intermediate state (State 3) is adopted, where pump wavelength tuning is performed concurrently with a gradual rotation of the pump polarization toward TE. The process ends in State 4, where a stable single soliton with improved power level is sustained by a purely TE-polarized pump. Through this dynamic pump polarization control, comb bandwidth is broadened as cooling power is efficiently utilized.

\subsection{Device design and passive characterization}
\begin{figure}[htpb]
  \centering
  \begin{minipage}{0.48\textwidth} 
    \centering
    \includegraphics[width=\linewidth]{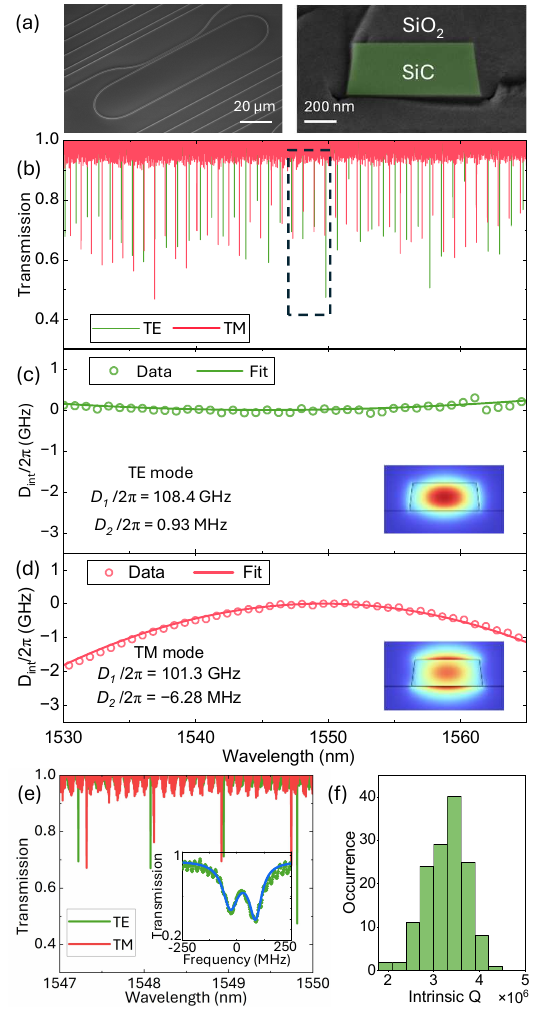}
    \caption{\textbf{Characterization of the fabricated SiC microresonator.} (a) SEM images of the fabricated SiC microresonator showing the top view (left) and cross section (right). (b) Measured transmission spectra of the fundamental TE₀₀ (green) and TM₀₀ (red) modes across the telecom C-band. (c), (d) Integrated dispersion of TE (anomalous) and TM (normal) modes, with insets showing simulated mode profiles. The small spatial overlap between the modes leads to negligible TE-TM coupling, consistent with the smooth dispersion and absence of avoided mode crossings near the pump wavelength. (e) Zoomed-in transmission spectra around the pump (corresponding to the dashed box in (b)), showing no significant extinction degradation. Lorentzian fitting of the closest TE resonance near the TE-TM mode crossing yields an intrinsic Q of $(3.5 \pm 0.07)\times 10^{6}$, close to the most probable value shown in the intrinsic Q histogram of TE mode (f), confirming that the high Q is preserved despite the nearby TE-TM resonance crossing.}
    \label{fig:2}
  \end{minipage}
\end{figure}

To implement the proposed thermal compensation scheme for robust soliton generation, precise engineering and characterization of the TE and TM mode families are essential. The racetrack microresonators were fabricated on a 430-nm-thick 4H SiC-on-insulator (SiCOI) sample using the optimized electron beam lithography \cite{Kim2023DesignPhotonics,Zheng2019High-quality-factorResonators} followed by reactive ion etching \cite{Zheng2019High-qualityCarbide-on-insulator}, the resulting sidewall angle is approximately 80\textdegree, as shown in the scanning electron microscope (SEM) images in Fig.~\ref{fig:2}(a). The waveguide geometry, with a top width of 1.05 \textmu m, is designed to firstly meet two key conditions: providing tailored dispersion for each mode family and suppressing strong TE-TM modal coupling \cite{Ye2023RobustMicroresonators}. Although the waveguide supports multiple spatial modes for each polarization, an Euler-bend racetrack design is employed to ensure effective single-mode operation of the microresonator \cite{Zheng2025EngineeredMicrocombs}. The bus-to-resonator coupling is realized using a symmetric directional coupler with a 440 nm gap and a 30 \textmu m  coupling length. The 1000-\textmu m-long cavity yields FSRs of approximately 108.4 GHz and 101.3 GHz for the fundamental TE and TM modes, respectively. The FSR difference ensures that a pair of TE and TM resonances exhibits the appropriate spectral separation required for thermal compensation. It also determines the spectral location of the resonance pair and can be engineered with sub-GHz precision by tailoring the waveguide dimensions. For instance, in our design, the FSR difference varies by less than 0.2 GHz and 0.1 GHz for ±5 \textdegree sidewall angle and ±30 nm waveguide width variations, respectively.

Fig.~\ref{fig:2}(b) presents the measured transmission spectra of the fundamental TE\textsubscript{00} (green) and TM\textsubscript{00} (red) modes. The integrated dispersion, defined as $D_{\text{int}}(\mu) = \omega_{\mu} - (\omega_{0} + D_{1}\mu)$, is extracted from the transmission spectra and plotted in Figs.~\ref{fig:2}(c) and~\ref{fig:2}(d). Here,  $\mu$ represents the mode index relative to the pump mode ($\mu = 0$), $\omega_{\mu}$ is the resonance frequency, and $D_{1}/2\pi$ denotes the FSR at the pump frequency. The TE mode is engineered to exhibit anomalous dispersion (second-order dispersion parameter is $D_2 / 2\pi = 0.93~\mathrm{MHz}\pm0.02~\mathrm{MHz}$), a precondition for the formation of bright Kerr solitons. Conversely, the TM mode is designed with large normal dispersion ($D_2 / 2\pi = -6.28~\mathrm{MHz}\pm0.07~\mathrm{MHz}$) to suppress optical parametric oscillation, ensuring the TM mode remains only for thermal compensation. All quoted uncertainties represent 95 \% confidence intervals derived from the covariance matrix of the polynomial fit to the measured resonance frequencies. Additionally, as illustrated by the simulated mode profiles in the insets of Figs.~\ref{fig:2}(c) and \ref{fig:2}(d), the high aspect ratio (width/height) of the waveguide leads to small spatial overlap between the TE and TM modes (calculated cross-polarization overlap of 0.02 \%), ensuring weak optical cross coupling. The experimental validation is evident in the magnified transmission spectra (Fig.~\ref{fig:2}(e)) and the smooth dispersion curves, where extinction ratios are not significantly perturbed and no avoided-mode crossing is observed in \textit{D\textsubscript{int}} around the pump wavelength ($\approx 1548$~nm), despite the close spectral proximity of the TE and TM resonances. The pronounced oscillations observed in the transmission background in Figs.~\ref{fig:2}(b) and~\ref{fig:2}(e) originate from facet reflections. The inset in Fig.~\ref{fig:2}(e) shows the fitting of a split TE resonance \cite{Li2016BackscatteringAnalysis} near the TE-TM crossing ($\approx 1549$~nm), showing an intrinsic $Q$ ($Q_{\text{int}}$) and loaded $Q$ ($Q_{\text{l}}$) of approximately $3.5 \times 10^{6}$ and $2.4 \times 10^{6}$, respectively. It aligns with the most probable intrinsic $Q$ factor as demonstrated in the histogram in Fig.~\ref{fig:2}(f) (measured from multiple resonances within the same resonator), confirming that the $Q$ factor is not degraded by the adjacent TM mode ($Q_{\text{int}}$ and $Q_{\text{l}}$ of approximately $1.27 \times 10^{6}$ and $0.31\times 10^{6}$, respectively) through cross coupling. The suppression of intermodal coupling not only preserves the high $Q$ factor, which is essential for achieving a low comb threshold, but also provides greater operation flexibility. Because the TE and TM modes can be brought into close spectral proximity without introducing mode-crossing-induced losses, efficient thermal compensation is achievable even at reduced pump powers.

 \begin{figure*}[htpb]
	\begin{center}
		\includegraphics[width=1\linewidth]{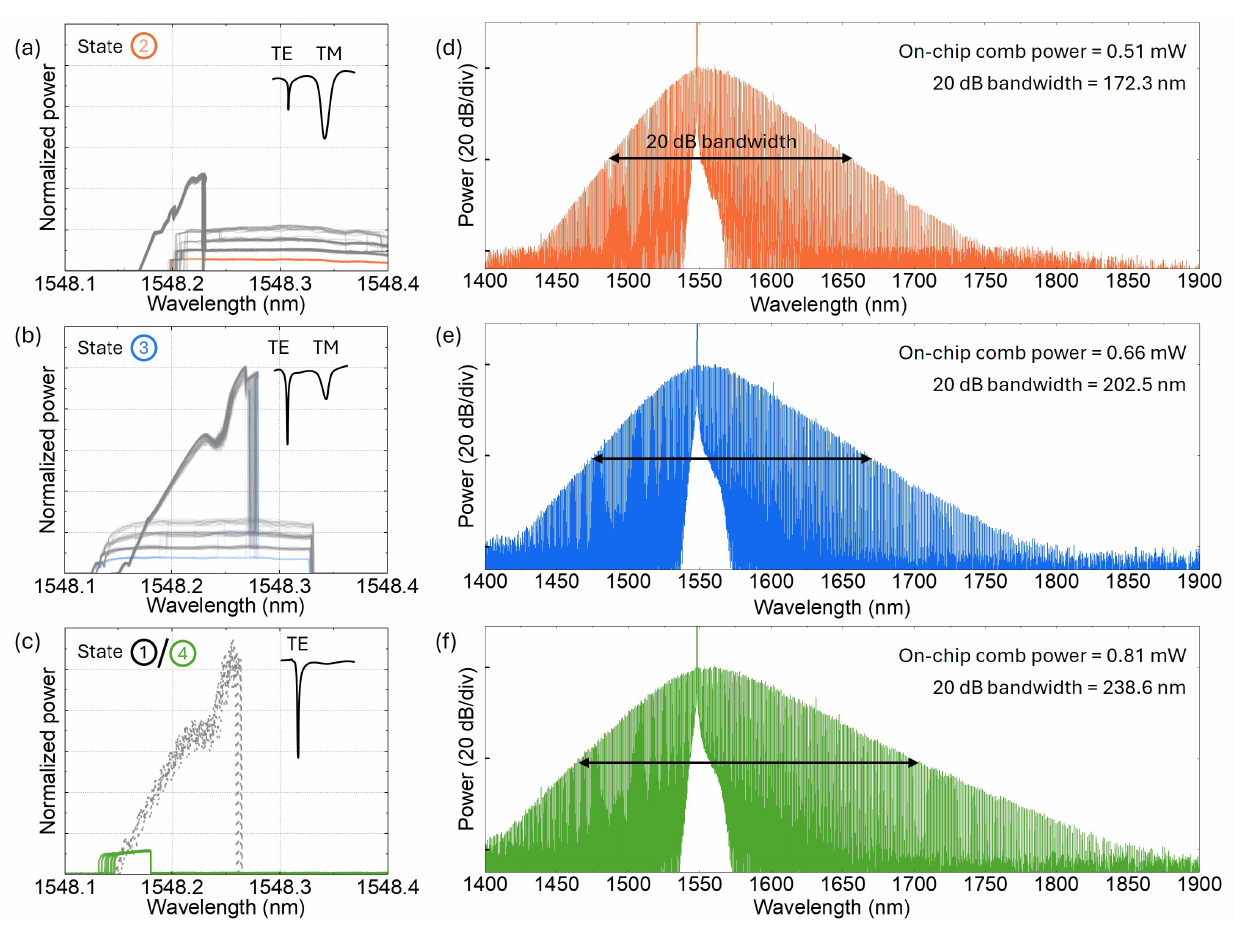}
		\caption{\textbf{Experimental demonstrations of soliton comb generation.} (a-c) Statistical comb power traces from 100 forward-backward pump-wavelength sweep cycles at a fixed on-chip pump power of 200 mW (insertion loss: 4 dB/facet), corresponding to State 2, State 3, and States 1 and 4, respectively. The trace opacity indicates the probability of accessing different soliton numbers, with colored segments highlighting the single-soliton detuning range. Insets show the transmission spectra of the pump resonances captured during the low-power characterization, revealing the mixed TE (green) and TM (red) mode composition of each state. In State 1 (c, dashed traces), no soliton step appears under a purely-TE polarized pump. High-power single-soliton operation is achieved by first using an obliquely polarized pump (a, b), followed by dynamic polarization rotation and pump wavelength tuning to reach State 4 (c, green traces). A larger TM component enhances single-soliton access probability and stability, making it easier for single-soliton initiation. (d-f), Measured single-soliton spectra at the maximum comb detuning for State 2 to State 4, showing a progressive bandwidth broadening as the power coupled to the TM cooling mode is redirected to the TE comb mode. A purely TE-polarized pump ultimately sustains a single-soliton comb spanning over 450 nm. }
		\label{fig:3}
	\end{center}
    \vspace{-10pt}
\end{figure*}   
   %Additionally, as illustrated by the simulated mode profiles in the insets of Figs.~\ref{fig:2}(c) and \ref{fig:2}(d), the high aspect ratio (width/height) of the waveguide leads to small spatial overlap between the TE and TM modes, ensuring weak optical cross coupling. The experimental validation is evident in the magnified transmission spectra (Fig. \ref{fig:2}(e)) and the smooth dispersion curves, where extinction ratios are not significantly perturbed and no avoided-mode crossing is observed in \textit{D\textsubscript{int}} around the pump wavelength ($ \approx $ 1548 nm), despite the close spectral proximity of the TE and TM resonances. The inset in Fig. 2(e) shows the fitting of a split TE resonance near the TE-TM crossing ($ \approx $ 1549 nm), showing that an intrinsic Q factor of approximately $3.5 \times 10^{6}$ is maintained. This value aligns with the most probable intrinsic Q factor as demonstrated in the histogram in Fig. \ref{fig:2}(f), confirming that the Q factor is not degraded by the adjacent TM mode through cross coupling.
   
   %The suppression of intermodal coupling not only preserves the high Q factor, which is essential for achieving a low comb threshold, but also provides greater operation flexibility. Because the TE and TM modes can be brought into close spectral proximity without introducing mode-crossing-induced losses, efficient thermal compensation is achievable even at reduced pump powers. 

\subsection{Soliton generation and cooling power reuse}
To investigate the thermal compensation mechanism, we perform experiments using both obliquely polarized and purely TE-polarized pumps near 1548 nm, where a TM resonance is close to the TE resonance. The TE resonance at the shorter wavelength (green) serves as the comb mode, while the TM resonance (red), separated by 4.5 GHz, functions as the cooling mode.

The experimental setup consists of a tunable continuous-wave (CW) laser amplified by an erbium-doped fiber amplifier (EDFA), with a band-pass filter employed to suppress amplified spontaneous emission (ASE) noise. Following polarization adjustment via a polarization controller (PC), the light is edge-coupled into the microresonator. The on-chip pump power is maintained at 200 mW, with an input coupling loss of 4 dB. The output light is split into three paths: the first is directed to an optical spectrum analyzer (OSA) for recording the comb spectra; the second is sent to a photodetector connected to an oscilloscope for monitoring the transmission; and the third passes through an optical filter for pump rejection, allowing a second photodetector to monitor the generated comb power. To ensure robustness, statistical analysis is performed over 100 consecutive pump frequency sweeps. The comb power traces are constructed by combining forward sweeps (capturing the MI and soliton steps) with backward sweeps (defining the full SER).

As indicated by the dashed traces in Fig.~\ref{fig:3}(c), conventional pumping with a purely TE-polarized pump (state 1) fails to produce a soliton step. Therefore, the TM fraction is adjusted to provide sufficient thermal compensation while maintaining enough TE power for soliton initiation. As shown in Figs.~\ref{fig:3}(a) and \ref{fig:3}(b), soliton steps corresponding to different soliton numbers are observed when the pump is obliquely polarized. In these plots, the opacity of each trace represents the probability of accessing different soliton states, while the colored segments highlight the detuning range of the single-soliton state. A larger TM fraction (state 2) creates the most favorable condition for single-soliton initiation, as evidenced by a significantly longer and more accessible (25 \% probability at state 2 compared to 8 \% at state 3) single-soliton step during forward pump wavelength tuning. In addition to proper polarization control, an appropriate pump power window is required for soliton initiation. Owing to the differential thermal responses between the comb and cooling modes, the pump power must be sufficient to thermally shift the comb-generating resonance to the cooling resonance, yet not so large that the thermo-optic effect drives it beyond the cooling resonance during thermal buildup. Consequently, the initial spectral separation between the two modes cannot be too large, otherwise the required thermal shift cannot be reached, nor too small, which would increase the risk of overshoot near the soliton threshold. We tested our device with TE-TM cold-cavity spectral separations ranging from 3.2 to 4.5 GHz, tuned via chip temperature between $25\,^\circ\mathrm{C}$ and $35\,^\circ\mathrm{C}$. Within this range, reliable soliton generation is achieved, with the minimum required pump power increasing from 32 to 126 mW.

Once the single soliton is established, the cooling power can be reused to support the comb by rotating the pump polarization toward TE and, if necessary, blue shifting the pump wavelength to preserve the soliton state. Following this procedure, a stable single-soliton state can subsequently be maintained using a purely TE-polarized pump, as shown in state 4 in Fig.~\ref{fig:3}(c). Notably, this dynamic control process does not impose strict requirements on the tuning speed. In this implementation, solitons were initiated with a~0.1 nm/s pump sweep, while the subsequent polarization and wavelength adjustment to state 4 was performed manually over a few seconds. The transition from state 2 to 4 involves a pump blue shift of approximately 0.08 nm, from roughly 1548.23 to 1548.15 nm. As long as the combined adjustment of pump detuning and polarization keeps the operating point within the evolving SER, the soliton remains stable. This flexibility allows the process to be fully automated by replacing the manual PC with a programmable electronic counterpart.
 
The single-soliton spectra presented in Figs.~\ref{fig:3}(d-f) are recorded at the pump wavelengths yielding the maximum comb detuning for each state, representing the condition for maximum achievable comb bandwidth. As the TE fraction increases, both the comb power and the 20-dB bandwidth are enhanced, owing to more pump power being coupled into the TE comb mode. Upon completing the cooling power reuse process to state 4, we achieve a purely TE-polarized single-soliton comb spanning more than 450 nm. The comb exhibits an on-chip power of 0.81 mW, a pump-to-comb conversion efficiency of 0.32 \%, and a 20 dB bandwidth of 238.6 nm, corresponding to increases of about 60 \% in both comb power and conversion efficiency, and about 39 \% in bandwidth compared with the obliquely polarized pump case in state 2 (0.51 mW, 0.20 \% conversion efficiency, and 172.3 nm).

The proposed dynamic polarization control scheme is not restricted to the present device and can be readily extended to other resonator geometries and wavelength regions \cite{Tan2025SinglePumping}, as it relies on general thermo-optic dynamics rather than specific material or structural parameters. For comb designs with high output power and conversion efficiency, typically enabled by larger dispersion and stronger bus-to-resonator coupling \cite{Jacobsen2025High-powerGeneration}, the resulting increase in soliton power helps mitigate thermo-optic instability, thereby facilitating the implementation of the proposed scheme. In contrast, for comb designs targeting low repetition rates (i.e., larger cavity length) or broadband operation (typically associated with smaller dispersion), thermo-optic instability becomes more pronounced. In such cases, stronger thermal compensation is required, which can be achieved by enhancing power buildup in the cooling mode, for example through higher $Q$ factors or operation closer to critical coupling.

More generally, the underlying principle of cooling power recovery is not restricted to dynamic polarization control. Active techniques, such as the phase modulation switch \cite{Shi2026AccessingSwitch}, represent alternative implementations, but with more complex experimental configurations. Regardless of the specific approach, effectively reusing the cooling power fundamentally increases the maximum achievable comb power and conversion efficiency.

\section{Conclusion}
In summary, we demonstrate thermally accessible single-soliton generation in SiC microresonators using an obliquely polarized pump, where the TE mode supports comb generation and the TM mode provides thermal compensation. To reuse the cooling power for enhancing the comb power and bandwidth after soliton initiation, a dynamic polarization rotation toward TE is implemented in combination with pump wavelength tuning to preserve the soliton state. By redirecting the cooling power to the comb-supporting TE mode, a single-soliton comb with a 108-GHz FSR and a spectral span exceeding 450 nm is experimentally achieved. This strategy effectively mitigates thermal instabilities for reliable soliton access without sacrificing cooling power, and is readily transferable to other nonlinear platforms, paving the way toward practical, broadband soliton microcombs for high-capacity optical communications, high-resolution spectroscopy, and optical frequency metrology. \bigskip\\
\textbf{Funding.}
This work is supported by Horizon Europe research and innovation programme under the Marie Skłodowska-Curie grant agreement No 101119968 (MicrocombSys), European Research Council (REFOCUS 853522), Independent Research Fund Denmark (ifGREEN 316400307A), Danish National Research Foundation (SPOC ref. DNRF123), and Innovationsfonden (GreenCOM 2079-00040B, EDOCS 4354-00020B). X. L. acknowledges the funding support from Maryland Industrial Partnerships (MIPS), in collaboration with LightSiNC NanoTech LLC, and the assistance from NGK Insulators LTD for the fabrication of SiC on insulator wafers. This fabrication was performed in part at the Cornell NanoScale Facility, an NNCI member supported by NSF Grant NNCI-1542081. \\
\smallskip\\
\textbf{Acknowledgment.} The authors thank Yuncong Liu and Jordan Stone for their valuable discussions of the manuscript. \\
\smallskip\\
\textbf{Disclosures.} The authors declare no conflicts of interest.\\
\smallskip\\
\textbf{Data Availability.} The data that support the findings of this study are available from the corresponding author upon reasonable request.
\smallskip\\
% \nocite{*}

\bibliographystyle{opticajnl}
\bibliography{references-haota}

\end{document}